\documentclass[aps,floatfix,showpacs,showkeys,amsmath]{revtex4}
\usepackage{color,graphicx}
\usepackage{dcolumn}
\usepackage{bm}
\usepackage{epsfig}
\usepackage{supertabular}
\usepackage[bookmarksopen]{hyperref}
\def\be {\begin{equation}}
\def\ee {\end{equation}}
\def\nn {\nonumber}
\def\bea {\begin{eqnarray}}
\def\eea {\end{eqnarray}}

\begin{document} 
\title{$\rho^0-\omega$ mixing in the presence of a weak magnetic field}
\bigskip
\bigskip
\author{Mahatsab Mandal$^{a,b}$}
\email{mahatsab@gmail.com}
\author{Arghya Mukherjee$^{a,d}$}
\email{arghya.mukherjee@saha.ac.in}
\author{Snigdha Ghosh$^{c,d}$}
\email{snigdha.physics@gmail.com}
\author{ Pradip Roy$^{b,d}$}
\email{pradipk.roy@saha.ac.in}
\author{Sourav Sarkar$^{c,d}$}
\email{sourav@vecc.gov.in}
\affiliation{$^a$Saha Institute of Nuclear Physics, 1/AF Bidhannagar
Kolkata - 700064, India}
\affiliation{$^b$Government General Degree College at Kalna-I, Burdwan - 713405, India}
\affiliation{$^c$Variable Energy Cyclotron Centre
1/AF Bidhannagar, Kolkata 700 064,
India}
\affiliation{$^d$Homi Bhabha National Institute, Training School Complex, Anushaktinagar, Mumbai - 400085, India}


\begin{abstract}
We calculate the momentum dependence of the $\rho^0-\omega$ mixing amplitude in vacuum with vector nucleon-nucleon 
interaction in presence of a constant homogeneous weak magnetic field background. The mixing amplitude is generated 
by the nucleon-nucleon ($NN$) interaction and thus driven by the neutron-proton mass difference along 
with a constant magnetic field. We find a significant effect of magnetic field on the mixing amplitude. 
We also calculate the Charge symmetry violating (CSV) $NN$ potential induced by the magnetic field 
dependent mixing amplitude. The presence of the magnetic field influences the $NN$ potential
substantially which can have important consequences in highly magnetized astrophysical compact objects, 
such as magnetars. The most important observation of this work is that the mixing amplitude is non-zero, 
leading to positive contribute to the CSV potential if the proton and neutron masses are taken to be equal.
\end{abstract}

\maketitle
\section{Introduction}
Recent years have witnessed significant progress in understanding the properties of strongly interacting nuclear matter in presence
of a magnetic background~\cite{lectnote871}. Such studies draw their motivation both from heavy-ion collision experiments 
and the physics of neutron stars. Magnetic field with the strength of $eB\sim(m_\pi^2-15m_\pi^2)$ can be achieved in the laboratory 
in non-central heavy-ion collisions at RHIC and LHC~\cite{nuclphy803,ijmp_24}. On the other hand, a similar environment can be 
expected in the interior of magnetars~\cite{AJ392,prl95,npb747,prd76,prl100,prl105,prd82}.
Several novel properties of the strongly interacting matter under extreme conditions have been 
studied like chiral magnetic effect~\cite{nuclphy797,nuclphy803,prd78,annphy_325}, magnetic catalysis~\cite{nuclphy462}, 
inverse magnetic catalysis~\cite{jhep1202}, phase structure of QCD~\cite{prd82}, superconductivity of vacuum~\cite{prd82_085011,
prl106,arxiv1208},  properties of mesons~\cite{prd86,plb_728,prd91,PhyRevD_93,prd94_09403,arxiv_1704,prd_sub}, photon polarization
~\cite{annphy_330,prd88}, dilepton production~\cite{prc88,prd94,annphy_376,arxiv_1704_01364} and many more.

Another phenomenologically important quantity to study concerns the charge symmetry of nuclear matter and its violation. 
Experimentally, charge symmetry violation (CSV) can be observed in a charge-conjugate system such as 
the difference between $pp$ and $nn$ scattering length in the $^1S_0$ state with the experimental value 
$\Delta a_{\rm CSV} = a^N_{pp}-a^N_{nn} =1.6 \pm 0.6 {~\rm fm/c}$~\cite{PhyRep194,plb_444,prl83}. 
Such a non-Coulombic interaction can also contribute to the binding energy difference of the light mirror 
nuclei  which is known as the Nolen-Schifer (NS) anomaly~\cite{annrevnucsci19,phylett11,plb277,prc63}. 
The CSV effect has been incorporated into the neutron-proton form-factor, the hadronic $\tau$ decay 
contribution~\cite{annrevnucsci56}, decay of the $\Psi^\prime\rightarrow (J/\Psi)\pi^0$, hadronic vacuum correction 
to $g-2$~\cite{prd83}, pion form factor~\cite{ppnp39}, and isospin asymmetric nuclear matter~\cite{plb339,prc63_015204,jpg35}. 
At the level of QCD, CSV occurs via the small mass difference between up and down quarks and via electromagnetic 
interaction of quarks~\cite{PhyRep194}.  Consequently, charge symmetry is violated at the hadronic level 
because of the neutron-proton mass difference. The major contribution to CSV is the isospin mixing 
of vector mesons, mainly $\rho^0-\omega$ mixing~\cite{npa249,npa287}, in single boson exchange model of the two 
nucleon force. Other examples of the mesons mixing are $\pi-\eta$ and $\pi-\eta^\prime$ mixing~\cite{prc26,prc48} 
the contribution of which is very small. The $\rho^0-\omega$ mixing is observed directly in the annihilation process 
$e^+e^-\rightarrow \pi^+\pi^-$ from which on-shell value of the mixing amplitude has been extracted from the experimental 
data at the $\omega$ pole and $<\rho^0|H|\omega> = -4520\pm600 {~\rm MeV^2}$~\cite{PhyRevC_36} is obtained. 
However, the mixing amplitude is not momentum independent in the $NN$ interaction, while the exchanged vector meson has 
a space-like four momentum. The $\rho^0-\omega$ mixing amplitude at the $\omega$ (or $\rho$) pole is quite different from 
its sign  and magnitude in the space-like region which is pertinent  to the construction of the CSV $NN$ potential. 
Goldman, Henderson, and Thomas~\cite{FBS_12} find that the $NN$ potential has a node at around 0.9 fm implying that 
the potential changes sign. Similar results were reported using several different theoretical approaches including 
mixing via $q\bar q$ loop driven by the $u-d$ quark mass difference~\cite{PhyLettB_317,PhyLettB_336}, and via $N\bar N$ loop 
using the small neutron-proton mass difference~\cite{PhyRevC_47}. Soon after their study it was argued in 
Ref.~\cite{PhyLettB_336} that the strong momentum dependent mixing amplitude must vanish at the transition from 
time-like to space-like region. Moreover, QCD sum-rule~\cite{PhyRevC_49}, calculation also gives 
a large momentum dependence of the coupling. Since the $NN$ potential involves the space-like
region, the long range $NN$ potential is strongly suppressed by the momentum dependent of $\rho-\omega$ mixing amplitude. 
As argued in Ref.~\cite{PhyRevC_52}, the off-shell dependence of $\rho^0-\omega$ mixing is not sufficient to determine 
the CSV potential. In contrast to the momentum dependent mixing amplitude, the "mixed propagator" field theory 
approach~\cite{Aip_412,Ppnp_39, WS_127} would restore the conventional role of the $\rho^0-\omega$ mixing.

It may further  be noted that in asymmetric nuclear matter $\rho^0-\omega$  mixing plays an important role in determining 
the symmetry energy which in turn affects the EOS of neutron star. 
It has been argued in Ref.~\cite{PhyRevC_80} that $\rho^0-\omega$ mixing has an important
effect on the symmetry energy. In fact the symmetry energy is softened both at sub- and super-saturation
densities. It is also to be noted that the change in symmetry energy modifies the equation of state (EOS)
of nuclear matter. Since the mixing depends both on the magnetic field and the density of the nuclear medium, there
$B$-dependent mixing in vacuum and intend to extend this calculation in nuclear matter in near future.
$\rho^0-\omega$ mixing in magnetic field might also affect the cooling of neutron 
star via neutrino emission through $NN\rightarrow NN\gamma\gamma$ where $NN$ cross section will be different because of the 
$B$-dependent $\rho^0-\omega$ mixing. In addition to that, the medium masses of $\rho$ and $\omega$  will also be affected in 
magnetic field due to $\rho^0-\omega$ mixing~\cite{PhyRevC_80}.

To explore the possible momentum dependence of the $\rho^0-\omega$ mixing amplitude  in the presence of a weak external 
magnetic field, we revisit the problem of $\rho^0-\omega$ mixing in vacuum. The mixing amplitude is generated by $N\bar N$ loop 
and led by the neutron-proton mass difference along with a background magnetic field. 
The effect of external magnetic field on 
fermionic propagators is taken into account using Schwinger propagator~\cite{PhyRep82}. In the present calculation, 
assuming that the magnetic field strength is weak i.e., $eB \ll m^2_{\rho/\omega}$, compatible with the strength observed in the 
interior of magnetars. In the presence of a magnetic field, the momentum dependence of $\rho^0-\omega$ mixing amplitude is modified, 
and it will affect the CSV $NN$ potential. Moreover, to examine the magnetic field dependent contribution, we also perform calculations
with equal nucleon masses in vacuum. 

The paper is organized as follows. In Sec. II, we discuss the formalism required for the explicit calculation of the momentum 
dependent $\rho^0-\omega$ mixing amplitude in presence of a weak magnetic field.  In Sec. III, we use the magnetic field dependent
mixing amplitude to determine the CSV $NN$ potential and discuss the numerical results.  Finally in Sec. IV we conclude with a brief
summary and discussions. Some details of the calculations are provided in the Appendix.

\section{$\rho^0-\omega$ meson mixing amplitude}
In the one-boson exchange (OBE) models, the $NN$ interaction is mediated by the exchange of several mesons.
For the purpose of this calculation, we are interested in the mixing between the neutral isovector 
$\rho^0$ meson and the isoscalar $\omega$ meson. The vector meson nucleon interaction Lagrangian corresponding to 
$\rho^0-\omega$ mixing that we use is the following:
\bea
\mathcal{L}_{\omega {\rm NN}} &=& g_\omega{\bar \Psi}\gamma_\mu\Phi^\mu_\omega\Psi,\\
\mathcal{L}_{\rho {\rm NN}} &=& g_\rho{\bar \Psi}\Big[\gamma_\nu+\frac{C_\rho}{2 M}\sigma_{\mu\nu}\partial^\mu\Big]  
\tau\cdot\Phi^\nu_\rho\Psi,
\eea
where $\Psi$ and $\Phi$ are the nucleon and meson fields, respectively. From the above interaction Lagrangian 
one can find the  vertex factors $\Gamma^{\mu}_{\omega}= g_\omega\gamma^\mu$ and 
${\tilde\Gamma}^\nu_\rho=g_\rho[\gamma^\nu+\frac{C_\rho}{2M}i\sigma^{\nu\lambda}q_\lambda]$.
In this paper we use the coupling constants determined by the Bonn group~\cite{PhyRep194}. 
The appropriate Bonn couplings are $g_\omega^2/4\pi = 10.6$, $g_\rho^2/4\pi=0.41$ and $C_\rho=f_\rho/g_\rho = 6.1$. 
In the present calculation, $NN\omega$ tensor coupling is not included for its negligible contribution.

\begin{figure}
\includegraphics[height=2.5cm, angle=0]{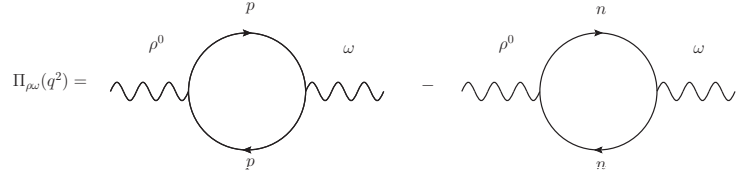}
\caption{ (Color online) Feynman diagram for $\rho^0-\omega$ mixing amplitude driven by the difference between proton and neutron loop.}
\label{fig1}
\end{figure}
The $\rho^0-\omega$ mixing amplitude is generated because of the difference between proton and neutron loop contribution as shown 
in Fig.\ref{fig1}:
\be
\Pi^{\mu\nu}_{\rho\omega}(q^2) =  \Pi^{\mu\nu(p)}_{\rho\omega}(q^2) - \Pi^{\mu\nu(n)}_{\rho\omega}(q^2),
\ee
where $p(n)$ stands for proton (neutron). 
The polarization tensor of $\rho^0-\omega$ mixing due to $NN$ excitations is calculated using standard Feynman rules and is given by
\be
i\Pi^{\mu\nu(N)}_{\rho\omega} (q^2) = \int\frac{d^4k}{(2\pi)^4} 
{\rm Tr}\Big[\Gamma^{\mu}_{\omega}(q)S_N(k){\tilde\Gamma}^\nu_\rho(-q) S_N(k+q)\Big],
\ee
where subscript N denotes either $p$ (proton) or $n$ (neutron). The Feynman propagator for the neutron is
\bea
S_n(k) = \frac{k\!\!\!/+m_n}{k^2-m_n^2}
\eea
To include the effect of a constant background magnetic field, we use Schwinger's proper time method~\cite{PhyRep82}. 
Without any loss of generality, we assume the magnetic field ${\bf B}$ along the $z$ direction. As we are interested in the 
weak field regime, i.e., $eB \ll m^2_{\rho/\omega}$, the magnetic field dependent proton propagator can be written as power 
series in $eB$, that up to order $(eB)^2$ read as~\cite{PhyRevD_62,PhyRevD_93} 
\bea
S_p(k) = S^{(0)}(k)+ S^{(1)}(k)+ S^{(2)}(k)
\eea
where
\bea
S^{(0)}(k) &=& \frac{k\!\!\!/+m_p}{k^2-m_p^2}\\
S^{(1)}(k) &=& eB\frac{i\gamma_1\gamma_2(\gamma\cdot k_{||}+m_p)}{(k^2-m_p^2)^2}\\
S^{(2)}(k) &=& (eB)^2\frac{-2k_\perp^2}{(k^2-m^2_p)^4}\Big[k\!\!\!/+m_p-\frac{\gamma\cdot k_\perp}{k^2_\perp}(k^2-m_p^2)\Big]
\eea
We decompose the metric tensor into two parts $g^{\mu\nu}=g^{\mu\nu}_{||}-g^{\mu\nu}_\perp$, 
where $g^{\mu\nu}_{||} = {\rm diag}(1,0,0,-1)$ and $g^{\mu\nu}_\perp = {\rm diag}(0,1,1,0)$. Also, 
we use $k_{||}^2 = k_0^2-k_3^2$ and $k_\perp^2 = k_1^2+k_2^2$.

The magnetic field independent vacuum contribution to the self-energy is 
\bea
i\Pi^{\mu\nu(N)}_{\rho\omega} (q^2) &=& \int\frac{d^4k}{(2\pi)^4} 
{\rm Tr}\Big[\Gamma^{\mu}_{\omega}(q)S_N(k){\tilde\Gamma}^\nu_\rho(-q) S_N(k+q)\Big]\nn\\
&=&{\rm g}_\omega {\rm g}_\rho \int \frac{d^4k}{(2\pi)^4}T^{\mu\nu}(k,k+q)\frac{1}{(k^2-m_N^{2}+i\epsilon)((k+q)^2-m_N^{2}+i\epsilon)}
\eea 
where 
\bea
{ T}^{\mu\nu}(k,k+q)&=&\Big(2k^\mu k^\nu + k^\mu q^\nu + k^\nu q^\mu-g^{\mu\nu}(k^2+k\cdot q-m_N^{2})
+\frac{C_\rho}{2 M}m_N(g^{\mu\nu}q^2-q^\mu q^\nu)\Big)
\eea
After the momentum integration, one may write the field free polarization tensor as 
\bea
\Pi^{\mu\nu(N)}_{\rho\omega\rm{(vac)}} (q^2) &=& (-g^{\mu\nu}+\frac{q^\mu q^\nu}{q^2})\Pi^{(N)}_{\rho\omega\rm{(vac)}} (q^2), 
\eea
where
\bea
\Pi^{(N)}_{\rho\omega\rm{(vac)}} (q^2) 
&=& -\frac{g_\rho g_\omega}{4\pi^2}q^2\int_0^1 dx \Bigg[2x(1-x)+\frac{C_\rho}{2 M}m_N\Bigg]
\Big(\frac{1}{\epsilon}-\gamma_E-\ln(\frac{\Delta}{\mu^2})\Big),
\eea
where $\Delta = m_N^{2}-x(1-x)q^2$, $\mu$ is an arbitrary renormalization scale.
$\gamma_E$ is the Euler-Mascheroni constant and $\epsilon=2-\frac{d}{2}$ contains the singularity, which diverges as $d\rightarrow4$. 
Since the individual self-energy contribution of proton and neutron diverges, the singularity can be removed by the difference 
between proton and neutron loop contribution and we obtain the magnetic field independent mixing amplitude as
\bea
\Pi_{\rho\omega\rm{(vac)}} (q^2) &=& \Pi^{(p)}_{\rho\omega\rm{(vac)}} (q^2)-\Pi^{(n)}_{\rho\omega\rm{(vac)}} (q^2)\nn\\
&=& \frac{g_\rho g_\omega}{4\pi^2}q^2\int_0^1 dx~ \Big(2x(1-x)+\frac{C_\rho}{2}\Big)
\ln\Big[\frac{m_p^{2}-x(1-x)q^2}{m_n^{2}-x(1-x)q^2}\Big]
\eea 
It can clearly be seen that if we do not distinguish between the proton and neutron mass, the mixing amplitude vanishes. 
In absence of magnetic field, the CSV $NN$ potential in vacuum does not exist for $m_p=m_n$. 

We now discuss the magnetic field dependent $\rho^0-\omega$ mixing amplitude. In this paper, we are mainly concerned with the $B$- 
dependent mixing amplitude up to $\mathcal{O}((eB)^2)$ which is reasonable in the weak field regime.
The first order contribution of magnetic field to $\rho^0-\omega$ mixing is (as explicitly shown in the Appendix A )
\bea
i\Pi^{\mu\nu1(p)}_{\rho\omega\rm{(vac)}} (q^2) &=& \int\frac{d^4k}{(2\pi)^4} 
{\rm Tr}\Big[\Gamma^{\mu}_{\omega}(q)S^{(0)}_p(k){\tilde\Gamma}^\nu_\rho(-q) S^{(1)}_p(k+q)
+\Gamma^{\mu}_{\omega}(q)S^{(1)}_p(k){\tilde\Gamma}^\nu_\rho(-q) S^{(0)}_p(k+q)\Big]\nn\\
 i\Pi^{\mu1(p)}_{~\mu,\rho\omega\rm{(vac)}} (q^2)
 &=&-8i \frac{C_\rho}{2 M} m_p
eB{\rm g}_\omega {\rm g}_\rho \int \frac{d^4k}{(2\pi)^4}
\epsilon^{\alpha\lambda\rho\sigma}k_\alpha q_\lambda  b_\rho u_\sigma \frac{1}{(k^2-m_p^2)((k+q)^2-m_p^2)^2}\nn\\
&=&0
\eea
Hence, the linear order contribution of order $eB$ vanishes.  

The second order contribution of magnetic field in $\rho^0-\omega$ mixing is given by(see Appendix B for details)
\bea
i\Pi^{\mu\nu2(p)}_{\rho\omega\rm{(vac)}} (q^2) &=& \int\frac{d^4k}{(2\pi)^4} 
{\rm Tr}\Big[\Gamma^{\mu}_{\omega}(q)S^{(2)}_p(k){\tilde\Gamma}^\nu_\rho(-q) S^{(0)}_p(k+q)
+\Gamma^{\mu}_{\omega}(q)S^{(0)}_p(k){\tilde\Gamma}^\nu_\rho(-q) S^{(2)}_p(k+q)\nn\\
&+& \Gamma^{\mu}_{\omega}(q)S^{(1)}_p(k){\tilde\Gamma}^\nu_\rho(-q) S^{(1)}_p(k+q)\Big]\nn\\
%
\Pi^{\mu2(p)}_{~\mu,\rho\omega\rm{(vac)}} (q^2) &=& (eB)^2\frac{{\rm g}_\omega {\rm g}_\rho}{\pi^2}\int_0^1 dx\,
\Bigg[x^3\,\Big[\frac{1}{\Delta}+\frac{x(1-x)q^2+x(4x-1)q^2_\perp+2m_p^2}{3\Delta^2}
+\frac{2x^2[x(1-x)q^2+2m_p^2]q^2_\perp}{3\Delta^3}\Big]\nn\\
&+& x^2\,\Big[\frac{1}{\Delta}-\frac{x(1-x)q^2_\perp}{\Delta^2}\Big]
-x(1-x)\Big[\frac{1}{2\Delta}+\frac{2x(1-x)q^2_{||}-m_p\frac{C_\rho}{2\bar M}(xq^2-(x+1)q^2_{||})}{4\Delta^2}\Big]\Bigg]\label{eb2}
\eea
It is clearly seen that the contribution of the magnetic field dependent mixing amplitude is finite; i.e., no divergences appear 
in the weak field limit. The correction term that is quadratic in field strength $eB$ contributes to  the $\rho^0-\omega$ mixing 
amplitude and we can express the magnetic field dependent part as 
$\Pi^{\rm eB}_{\rho\omega\rm{(vac)}} = -\frac{1}{3} \Pi^{\mu2(p)}_{~\mu,\rho\omega\rm{(vac)}}$. 
In presence of the external magnetic field, the total contribution to the mixing amplitude can be written as
\bea
\Pi_{\rho\omega}^{\rm t}(q^2) &=& \Pi_{\rho\omega\rm{(vac)}} (q^2)+ \Pi^{\rm eB}_{\rho\omega\rm{(vac)}}(q^2)
\eea

\begin{figure}
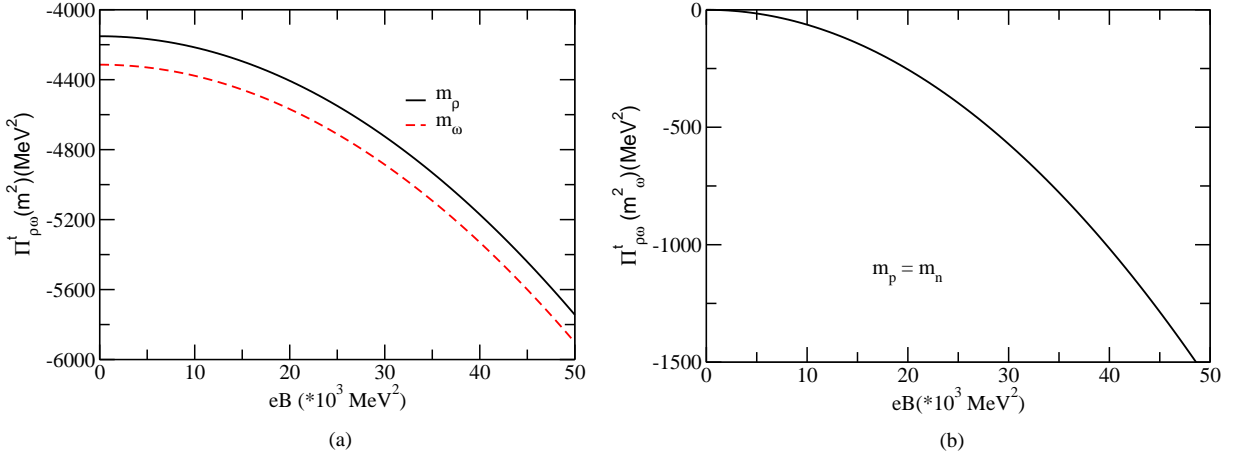

 \includegraphics[height=6cm, angle=0]{Bmass.eps}~~~\includegraphics[height=6cm, angle=0]{B_0_momega.eps}
\caption{ (Color online) Mixing amplitude as a function of the magnetic field.}
\label{fig2}
\end{figure}

In absence of magnetic field, we obtain the mixing amplitude at the on-shell $\omega$ and $\rho$ meson point 
$\Pi_{\rho\omega}(m^2_\omega) = -4314~\rm{MeV}^2$ and  $\Pi_{\rho\omega}(m^2_\rho) = -4152~\rm{MeV}^2$ respectively, 
which compares well with the experimental values~\cite{PhyRevC_36}.  In Fig.~\ref{fig2}(a) we have shown the variation 
of the mixing amplitude at the point $(q^2=m^2_{\rho/\omega})$ with weak external magnetic field. We have used the 
condition that the strength of the external field is much lower than the square of the vector meson mass, 
i.e., $eB \ll m^2_{\rho/\omega}$. In both the meson mass, we have observed that the the mixing amplitude, 
$\Pi_{\rho\omega}(q^2=m^2_{\rho/\omega})$ decreases with the increase of external magnetic field strength.  
In presence of background magnetic field, the mixing amplitude is non-zero, even in the limit $m_p=m_n$ 
as shown in Fig.~\ref{fig2}(b). It is seen that, taking the limit ($m_p=m_n$), the mixing amplitude 
vanishes at $eB =0$ and hence, we see a decreasing behavior of mixing amplitude with increasing $eB$.  
\begin{figure}
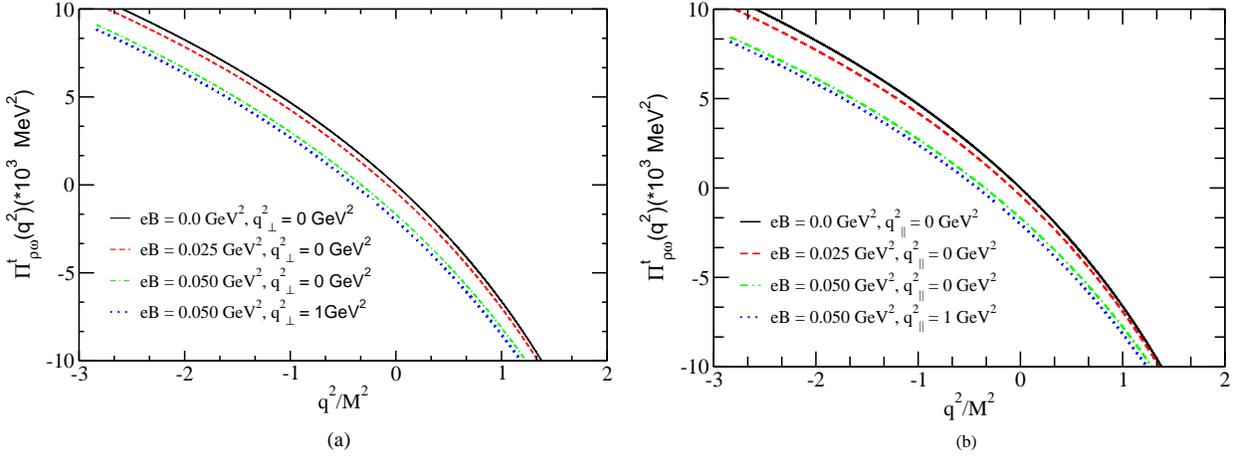

 \includegraphics[height=6cm, angle=0]{amp1.eps}~~~\includegraphics[height=6cm, angle=0]{amp2.eps}
\caption{ (Color online) $\rho^0-\omega$ mixing amplitude as a function of $q^2$ with different values of $eB$ in weak field regime.
The left (right) panel also shows the variation of mixing amplitude for two different values of $q_{\perp}^2$ ($q_{||}^2$).}
\label{fig3}
\end{figure}

The momentum dependence of the $\rho^0-\omega$ mixing amplitude is displayed in Fig.~\ref{fig3} at a different magnetic 
field strength. In absence of $eB$, the mixing amplitude has a node at exactly $q^2=0$~\cite{PhyRevC_47,PhyLettB_336} 
and, consequently, there is a change of sign of the mixing amplitude.  Fig.~\ref{fig3}(a) displays the mixing amplitude 
which is diminished with increasing values of $eB$ at same values of $q^2_\perp$. 
It is also clearly noticed that the value of $\Pi_{\rho\omega}^{\rm t}$ decreases with the increase of $q^2_\perp$ at fixed values of 
background magnetic field. Similar behavior can be observed in Fig.~\ref{fig3}(b) where $eB$ is varied keeping  $q^2_{||}$ fixed.
The effect of magnetic field on the mixing amplitude is greater in the time-like region than the space-like region. It is clearly
visible that the node is shifted towards the space-like region in presence of magnetic field.     

\section{Charge symmetry violating potential}
Now we will evaluate the CSV $NN$ potential induced by the $\rho^0-\omega$ mixing in presence of an external weak magnetic field. 
The momentum space CSV potential due to $\rho^0-\omega$ mixing is given by~\cite{PhyRevC_47,PhyRevC_78}:
\be
V^{NN}_{\rho\omega}({\bf q}) = -\frac{{\rm g}_\omega {\rm g}_\rho \Pi_{\rho\omega}^{\rm t}({\bf q})}
{({\bf q}^2+m^2_\rho)({\bf q}^2+m^2_\omega)}
\label{CSV_1}
\ee
Here, we neglected the contribution due to the external legs. 
Because of the extended structure of hadrons, one needs to incorporate meson-nucleon vertex correction which would be 
sufficient to take into account the inner structure of the hadrons. In our analysis, form factors are introduced by parameterizing
the point coupling as~\cite{PhyRep194}:
\be
{\rm g}_i\rightarrow {\rm g}_i\Big(\frac{\Lambda_i^2-m^2_i}{\Lambda_i^2+{\bf q}^2}\Big)
\ee
The cutoff parameter $\Lambda_i$ can be related directly to the hadron size and the numerical values for the cutoffs ($\Lambda_i$)
are determined from the fit of the empirical $NN$ data~\cite{PhyRep194}.

To convert the CSV potential to configuration space, we make use of the identity
\be
\frac{1}{({\bf q}^2+m^2_\rho)({\bf q}^2+m^2_\omega)} = \frac{1}{m^2_\omega-m^2_\rho}\Big(\frac{1}{{\bf q}^2+m^2_\rho}
-\frac{1}{{\bf q}^2+m^2_\omega}\Big),
\ee
and find the CSV potential with the on-shell mixing amplitude in coordinate space through the Fourier transformation of Eq.~\ref{CSV_1}.
This yields the result
\bea
V^{NN}_{\rho\omega}(r) = - \frac{{\rm g}_\omega {\rm g}_\rho}{4\pi}\frac{ \Pi_{\rho\omega}^{\rm t}({m^2_\omega})}
{m^2_\omega-m^2_\rho}\Big(\frac{e^{-m_\rho r}}{r}-\frac{e^{-m_\omega r}}{r}\Big)
\label{potential1}
\eea
With the inclusion of form factors the CSV potential reduces to 
\bea
V^{NN}_{\rho\omega}(r) &=& - \frac{{\rm g}_\omega {\rm g}_\rho}{4\pi}\frac{ \Pi_{\rho\omega}^{\rm t}({m^2_\omega})}{m^2_\omega-m^2_\rho}
\Bigg[\frac{\Lambda^2_\omega-m^2_\omega}{\Lambda^2_\omega-m^2_\rho} \frac{e^{-m_\rho r}}{r}-
\frac{\Lambda^2_\rho-m^2_\rho}{\Lambda^2_\rho-m^2_\omega} \frac{e^{-m_\omega r}}{r}\nn\\
&+&\frac{m^2_\omega-m^2_\rho}{\Lambda^2_\omega-\Lambda^2_\rho}\Big(\frac{\Lambda^2_\omega-m^2_\omega}{\Lambda^2_\rho-m^2_\omega}
\frac{e^{-\Lambda_\rho r}}{r}-\frac{\Lambda^2_\rho-m^2_\rho}{\Lambda^2_\omega-m^2_\rho}
\frac{e^{-\Lambda_\omega r}}{r}
\Big)\Bigg]
\label{potential2}
\eea 
It is to be noted that in the limit $\Lambda_i\rightarrow \infty$, Eq.~\ref{potential2} reduces to Eq.~\ref{potential1}. 

\begin{figure}
 \includegraphics[height=6cm, angle=0]{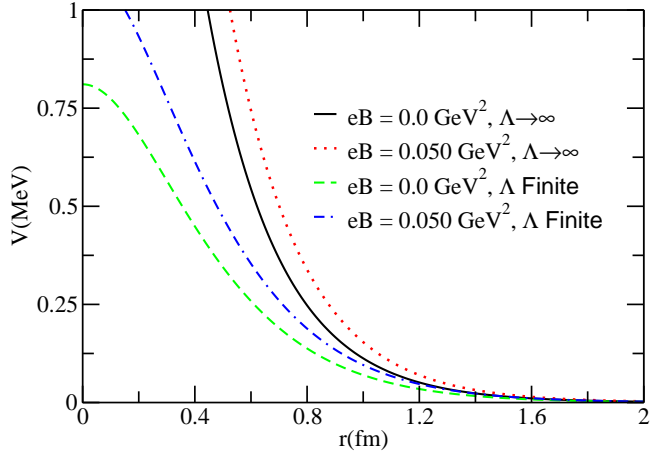}
\caption{ (Color online) The contribution from $\rho^0-\omega$ mixing to the $NN$ potential as a function of $NN$ 
separation using the on-shell value for the mixing amplitude in different magnetic fields.}
\label{fig4}
\end{figure}
In Fig.~\ref{fig4} we show the contribution to the $NN$ potential in configuration space with the constant on-shell 
mixing amplitude. We see that there is a stronger suppression of the $NN$ potential going from the point coupling to the 
form factor. Magnetic field dependent mixing amplitude leads to a clear enhancement of the $NN$ potential compared with 
the magnetic field independent on-shell mixing amplitude. 

\begin{figure}
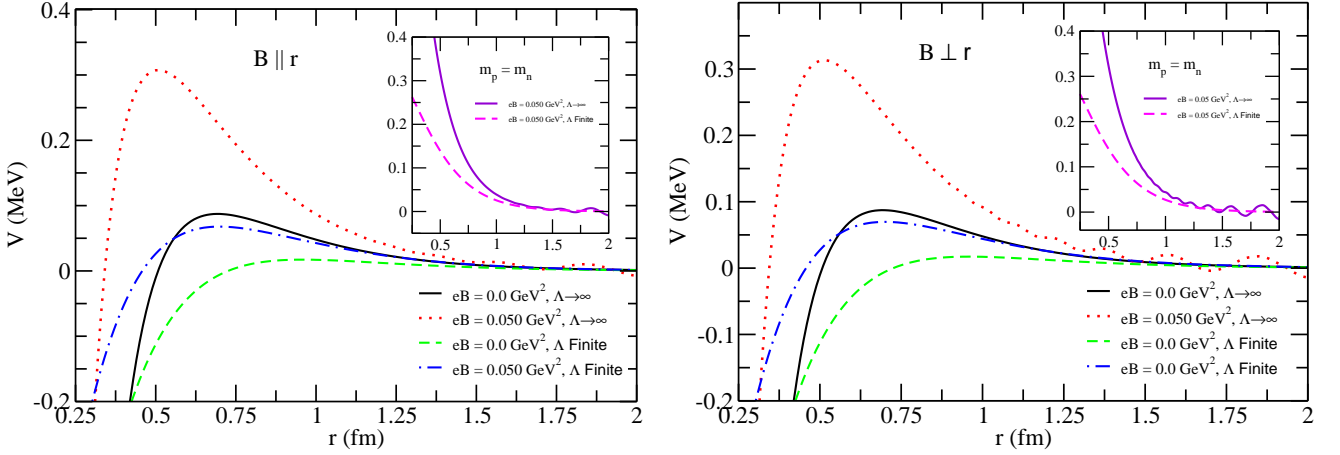

 \includegraphics[height=6cm, angle=0]{pot1.eps}~~~\includegraphics[height=6cm, angle=0]{pot2.eps}
\caption{ (Color online) The off-shell contribution from $\rho^0-\omega$ mixing to the CVS potential as a function of $NN$ 
separation. Both cases without magnetic field ($eB = 0$) and with magnetic field ($eB = 0.05~{\rm GeV}^2$)  are shown. 
Left panel: $B || r$. Right panel $B \perp r$. The CSV potential for $m_p = m -n$ is shown in the inset}
\label{fig5}
\end{figure}

As we have already mentioned the $\rho^0-\omega$ mixing amplitude is strongly dependent on momentum. Here, we calculate the contribution
of the off-shell dependence of $\rho^0-\omega$ mixing in the CSV potential. Magnetic field independent CSV potential can be obtained 
analytically~\cite{PhyRevC_78} but in case of non-zero $eB$ we discuss the numerical results. We solve the magnetic field dependent
CSV potential for two special cases: (a) $\bf B||r$ and (b) $\bf B\perp r$. In Fig.~\ref{fig5} we present the role of the off-shell
contribution of $\rho^0-\omega$ mixing in the CSV $NN$ potential. The contribution of the background magnetic field 
to the $NN$ potential is clearly shown in both the graphs. We see that the $B$-independent CSV potential have a node around 
$0.9$ fm ~\cite{FBS_12,PhyRevC_47} with form factors. As the magnetic field is turned on, the occurrence of the node in 
the potential is around at $0.35$ fm (at $eB = 0.05~{\rm GeV}^2$). We also notice that a non-zero $B$-dependent $\rho^0-\omega$ 
contribution to the $NN$ interaction is found to be much larger than without $B$-dependent mixing amplitude. It is also interesting 
to examine the CSV potential in presence of weak field regime at $m_p=m_n$, as is shown in the inset of Fig.~\ref{fig5}. 
It is seen that the effect of magnetic field on the $NN$ potential is found to be always positive in space-like region, and 
consequently, there is no node in the $NN$ potential which leads to a significant effect on CSV.

\section{Summary and Conclusion}
In the present paper, we have investigated the momentum dependence of $\rho^0-\omega$ mixing amplitude as well as the role of 
momentum dependence of $\rho^0-\omega$ mixing amplitude in CSV $NN$ potential in the presence of an external magnetic field for 
the first time. The $\rho^0-\omega$ mixing was assumed to be generated by the $NN$ loops and hence driven by the neutron-proton 
mass difference along with a constant magnetic field. We have restricted ourselves to the weak field limit, where the external 
field satisfies $eB\ll m^2_{\rho/\omega}$ and used the Schwinger's proper-time method to describe the fermionic propagator. 
The effect of the background magnetic field appears as a correction to the momentum dependence of $\rho^0-\omega$ mixing amplitude, 
which is relevant to study the properties of magnetars and magnetized hadronic medium relativistic heavy-ion collisions. 
Although in the weak field limit, the first correction is quadratic in the field. One has to also take into account the 
linear order corrected fermionic propagator in $B$. We find that the presence of the 
magnetic field modifies the mixing amplitude. It is seen that the mixing amplitude decreases with the increase of the 
strength of the magnetic field at the on-shell meson mixing point. This happens even if the Hamiltonian preserves the isospin
symmetry, i.e., $m_p=m_n$. It is important to note that the change in 
the sign of the momentum dependence of $\rho^0-\omega$ mixing amplitude is shifted towards the space-like region for non-zero 
$eB$ in contrast to the result found in the absence of magnetic field. Furthermore, the $NN$ potential generated by the off-shell 
dependence of $\rho^0-\omega$ mixing is evaluated numerically. We have found that a node in the $NN$ potential occurrs at 
$r\,\sim\,0.35$ fm for $eB = 0.05~{\rm GeV}^2$. Interestingly, we also find that the effect of the magnetic field to the $NN$ 
potential is always positive in the space-like region if we assume that each of the nucleon masses are taken to be equal.  
Moreover, one needs to extend this calculation in the dense medium to study the changes in various properties of magnetars.

\section*{Appendix A: Calculation of $\Pi^{\mu1(p)}_{~\mu,\rho\omega\rm{(vac)}}$}
We have 
\bea
i\Pi^{\mu\nu1(p)}_{\rho\omega\rm{(vac)}} = \int \frac{d^4k}{(2\pi)^4}\, eB\, {\rm g}_\omega {\rm g}_\rho\Big[
\frac{T_1^{\mu\nu1}}{(k-m_p^2)((k+q)^2-m_p^2)^2}+\frac{T_2^{\mu\nu1}}{(k-m_p^2)^2((k+q)^2-m_p^2)}\Big]
\eea
where 
\bea
T_1^{\mu\nu1} &=& {\rm Tr}[\gamma^\mu(k\!\!\!/+m_p)(\gamma^\nu-\frac{C_\rho}{2\bar M}i\sigma^{\nu\lambda}q_\lambda)
i\gamma_1\gamma_2(\gamma\cdot (k+q)_{||}+m_p)],\nn\\
T_1^{\mu\nu2} &=& {\rm Tr}[\gamma^\mu i\gamma_1\gamma_2(\gamma\cdot k_{||}+m_p)
(\gamma^\nu-\frac{C_\rho}{2\bar M}i\sigma^{\nu\lambda}q_\lambda)((k\!\!\!/+q\!\!\!/)+m_p)
\eea
We use $i\gamma_1\gamma_2 = -\gamma^5 b\!\!\!/u\!\!\!/$, with $u^{\mu} = (1,0,0,0)$ and $b^\mu = (0,0,0,1)$.
Using that 
\bea
T^{\mu1}_{~\mu1} &=& -4i \frac{C_\rho}{2\bar M} m_p \epsilon^{\alpha\lambda\rho\sigma}k_\alpha q_\lambda b_\rho u_\sigma \nn\\
T^{\mu1}_{~\mu2} &=& -4i \frac{C_\rho}{2\bar M} m_p \epsilon^{\alpha\lambda\rho\sigma}(k+q)_\alpha q_\lambda b_\rho u_\sigma \nn\\
\eea
Therefore, we can write the linear order contribution of magnetic field in the $\rho^0-\omega$ mixing amplitude
\bea
 i\Pi^{\mu1(p)}_{~\mu,\rho\omega\rm{(vac)}} (q^2)
 &=&-8i \frac{C_\rho}{2 {\bar M}} m_p
\,eB\,{\rm g}_\omega {\rm g}_\rho \int \frac{d^4k}{(2\pi)^4}
\epsilon^{\alpha\lambda\rho\sigma}k_\alpha q_\lambda  b_\rho u_\sigma \frac{1}{(k^2-m_p^2)((k+q)^2-m_p^2)^2}\nn\\
&=& -8i \frac{C_\rho}{2 {\bar M}} m_p
\,eB\,{\rm g}_\omega {\rm g}_\rho \int_0^1 dx \, 2x \int \frac{d^4k}{(2\pi)^4}
\epsilon^{\alpha\lambda\rho\sigma} q_\lambda  b_\rho u_\sigma\frac{(k-xq)_\alpha}{[k^2-\Delta]^3}\nn\\
&=&0
\eea
Here, the integration involving linear terms in $k$ is zero and $\epsilon^{\alpha\lambda\rho\sigma} q_\alpha q_\lambda =0 $ due to 
the antisymmetric properties of Levi-Civita tensor.

\section*{Appendix B: Calculation of $\Pi^{\mu2(p)}_{~\mu,\rho\omega\rm{(vac)}}$}
We have 
\bea
i\Pi^{\mu\nu2(p)}_{\rho\omega\rm{(vac)}} &=& \int \frac{d^4k}{(2\pi)^4}\, (eB)^2\, {\rm g}_\omega {\rm g}_\rho\Big[
T_1^{\mu\nu2}\frac{-2k^2_\perp}{(k^2-m_p^2)^4((k+q)^2-m_p^2)}+
T_2^{\mu\nu2} \frac{-2(k+q)^2_\perp}{(k^2-m_p^2)((k+q)^2-m_p^2)^4}\nn\\&+&
T_3^{\mu\nu2}\frac{1}{(k^2-m_p^2)^2((k+q)^2-m_p^2)^2}
\Big]
\eea
where
\bea
T_1^{\mu\nu2}&=&{\rm Tr}[\gamma^\mu(k\!\!\!/+m_p-\frac{\gamma\cdot k_\perp}{k_\perp^2}(k^2-m_p^2))
(\gamma^\nu-\frac{C_\rho}{2\bar M}i\sigma^{\nu\lambda}q_\lambda)
(k\!\!\!/+q\!\!\!/+m_p)],\nn\\
&=& 4\Big[k^\mu p^\nu+p^\mu k^\nu-g^{\mu\nu}(k\cdot p)-\frac{k^2-m^2_p}{k^2_\perp}
(k_\perp^\mu p^\nu+p^\mu k_\perp^\nu-g^{\mu\nu}(k_\perp\cdot p))+g^{\mu\nu}m^2_p \nn\\
&-& m_p\frac{C_\rho}{2\bar M}[q^\mu p^\nu -g^{\mu\nu}(p\cdot q)+g^{\mu\nu}(k\cdot q)-q^\mu k^\nu
-\frac{k^2-m^2_p}{k^2_\perp}(g^{\mu\nu}(q\cdot k_\perp)-q^\mu k^\nu_\perp)] \Big]
\eea
\bea
T^{\mu\nu2}_2 &=&{\rm Tr}[\gamma^\mu(k\!\!\!/+m_p)(\gamma^\nu-\frac{C_\rho}{2\bar M}i\sigma^{\nu\lambda}q_\lambda)
(k\!\!\!/+q\!\!\!/+m_p-\frac{\gamma\cdot (k+q)_\perp}{(k+q)_\perp^2}((k+q)^2-m_p^2))],\nn\\
&=&4\Big[k^\mu p^\nu+p^\mu k^\nu-g^{\mu\nu}(k\cdot p)-\frac{p^2-m^2_p}{p^2_\perp}
(k^\mu p_\perp^\nu+p_\perp^\mu k^\nu-g^{\mu\nu}(k\cdot p_\perp))+g^{\mu\nu}m^2_p \nn\\
&+& m_p\frac{C_\rho}{2\bar M}[q^\mu k^\nu -g^{\mu\nu}(k\cdot q)+g^{\mu\nu}(p\cdot q)-q^\mu p^\nu
-\frac{p^2-m^2_p}{p^2_\perp}(g^{\mu\nu}(q\cdot p_\perp)-q^\mu p^\nu_\perp)] \Big]\nn\\
\eea
where $p=k+q$. Now, we replace $k\leftrightarrow k+q$ and we find
\bea
T^{\mu\nu2}_2  &=& 4\Big[k^\mu p^\nu+p^\mu k^\nu-g^{\mu\nu}(k\cdot p)-\frac{k^2-m^2_p}{k^2_\perp}
(k_\perp^\mu p^\nu+p^\mu k_\perp^\nu-g^{\mu\nu}(k_\perp\cdot p))+g^{\mu\nu}m^2_p \nn\\
&+& m_p\frac{C_\rho}{2\bar M}[q^\mu p^\nu -g^{\mu\nu}(p\cdot q)+g^{\mu\nu}(k\cdot q)-q^\mu k^\nu
-\frac{k^2-m^2_p}{k^2_\perp}(g^{\mu\nu}(q\cdot k_\perp)-q^\mu k^\nu_\perp)] \Big]
\eea
and 
\bea
T^{\mu\nu2}_3 &=& {\rm Tr}[\gamma^\mu i\gamma_1\gamma_2(\gamma\cdot k_{||}+m_p)
(\gamma^\nu-\frac{C_\rho}{2\bar M}i\sigma^{\nu\lambda}q_\lambda) i\gamma_1\gamma_2(\gamma\cdot (k+q)_{||}+m_p)]\nn\\
\eea
The contribution of the magnetic field comes from the $\mathcal O((eB)^2)$ terms:
\bea
i\Pi^{\mu2(p)}_{~\mu,\rho\omega\rm{(vac)}} &=& \int \frac{d^4k}{(2\pi)^4}\, (eB)^2\, {\rm g}_\omega {\rm g}_\rho\Bigg[ 
32\Big[\frac{(k^2+k\cdot q-2m_p^2)k^2_\perp}{(k^2-m_p^2)^4((k+q)^2-m_p^2)}
+\frac{k^2_\perp+k_\perp\cdot q_\perp}{(k^2-m_p^2)^3((k+q)^2-m_p^2)}\Big]\nn\\&+&
\frac{8 k\cdot (k+q)_{||} + 
4 m_p\frac{C_\rho}{2\bar M}(k_{||}\cdot q_{||}-k\cdot q -q_{||}^2)}{(k^2-m_p^2)^2((k+q)^2-m_p^2)^2}
\Bigg]\label{appendix1}
\eea
Using the standard procedure of Feynman parametrization and evaluation of the momentum integral and Eq.~\ref{appendix1} 
reduce to Eq.~\ref{eb2}


\begin{thebibliography}{50}
 \bibitem{lectnote871}D.~E.~Kharzeev,~K.~Landsteiner,~A.~Schmitt and H.~U.~Yee,~Lect.~Notes Phys.~\textbf{871},~1~(2013).

\bibitem{nuclphy803} D.~E.~Kharzeev,~L.~D.~McLerran and H.~J.~Warringa,~~Nucl.Phys. A \textbf{803},227 (2008).
\bibitem{ijmp_24} V. Skokov, A. Y. Illarionov, and V. Toneev, Int. J. Mod. Phys. A {\bf24}, 5925 (2009).

\bibitem{AJ392} R. C. Duncan and C. Thompson, Astrophys. J. {\bf392}, L9 (1992)
\bibitem{prl95} E. J. Ferrer, V. de la Incera and C. Manuel, Phys. Rev. Lett. {\bf95}, 152002 (2005). 
\bibitem{npb747} E. J. Ferrer, V. de la Incera and C. Manuel, Nucl. Phys. {\bf B747}, 88 (2006).
\bibitem{prd76} E. J. Ferrer and V. de la Incera, Phys. Rev. D {\bf76}, 045011 (2007).
\bibitem{prl100} K. Fukushima and H. J. Warringa, Phys. Rev. Lett. {\bf100}, 032007 (2008).
\bibitem{prl105} B. Feng, D. Hou, H. c. Ren and P. P. Wu, Phys. Rev. Lett. {\bf105}, 042001 (2010).
\bibitem{prd82} S. Fayazbakhsh and N. Sadooghi, Phys. Rev. D 82, 045010 (2010).

\bibitem{nuclphy797} D.~E.~Kharzeev, and A.~Zhitnitsky,~Nucl.Phys. A \textbf{797},67 (2007). 

\bibitem{prd78} K.~Fukushima,~D.~E.~Kharzeev and H.~J.~Warringa,Phys.Rev.~D \textbf{78},~074033 (2008).
\bibitem{annphy_325} D.~E.~Kharzeev, Ann.~Phys. (N.Y)~{\bf 325}, 205 (2010).  

\bibitem{nuclphy462}V.~P.~Gusynin,~V.~A.~Miransky and I.~A.~Shovkovy, Nucl.Phys.  \textbf{B462}, 249 (1996), 
Nucl.Phys.  \textbf{B563}, 361 (1999).

\bibitem{jhep1202}G.~S.~Bali,~F.~Bruckmann,~G.~Endrodi,~Z.~Fodor,~S.~D.~Katz,~S.~Kreig,~A.~Schafer and K.~K.~Szabo,
JHEP \textbf{1202},~044 (2012).



\bibitem{prd82_085011} M.~N.~Chernodub,~Phys.~Rev.~D \textbf{82},~085011 (2010). 
\bibitem{prl106}M.~N.~Chernodub,~Phys.~Rev.~Lett. \textbf{106}, 142003 (2011).
\bibitem{arxiv1208} M.~N.~Chernodub,~ Lect.~Notes Phys. ~\textbf{871},~143 (2013).

\bibitem{prd86} J. O. Andersen, Phys. Rev. D {\bf86} 025020 (2012).
\bibitem{plb_728} G. Colucci, E. S. Fraga, and A. Sedrakian, Phys. Lett. {\bf B728} 19 (2014).
\bibitem{prd91} H. Liu, L. Yu, and M. Huang, Phys. Rev. D {\bf91}, 014017 (2015).
\bibitem{PhyRevD_93} S. P. Adhya, M. Mandal, S. Biswas, and P. K. Roy, Phys. Rev. D {\bf93}, 074033 (2016). 
\bibitem{prd94_09403} S. Ghosh, A. Mukherjee, M. Mandal, S. Sarkar, and P. Roy, Phys. Rev. D {\bf94}, 094043 (2016).
\bibitem{arxiv_1704} S. Ghosh, A. Mukherjee, M. Mandal, S. Sarkar, and P. Roy, arXiv:1704.05319 [hep-ph].
\bibitem{prd_sub} A. Mukherjee, S. Ghosh, M. Mandal, S. Sarkar, and P. Roy, Submitted in Phys. Rev. D.

\bibitem{annphy_330} K. Hattori and K. Itakura, Ann. Phys. (Amsterdam) {\bf 330} 23 (2013), {\bf 224}, 58 (2013).
\bibitem{prd88} F. Karbstein  Phys. Rev. D {\bf 88} 085033 (2013).

\bibitem{prc88} K. Tuchin, Phys. Rev. C {\bf88} 024910 (2013).
\bibitem{prd94} A. Bandyopadhyay, C. A. Islam and M. G. Mustafa, Phys. Rev. D {\bf 94}, 114034 (2016).
\bibitem{annphy_376} N. Sadooghi and F. Taghinavaz, Annals Phys. 376, 218 (2017).
\bibitem{arxiv_1704_01364} A. Bandyopadhyay and S. Mallik, arXiv:1704.01364 [hep-ph].

\bibitem{PhyRep194} G. A. Miller, M. K. Nefkens, and I. Slaus, Phys. Rep. {\bf194}, 1 (1990).
\bibitem{plb_444} C. R. Howell {\it et al}., Phys. Lett {\bf B444}, 252 (1998).
\bibitem{prl83} D. E. Gonzalez Trotter {\it et al}., Phys.Rev.  Lett {\bf 83},3788 (1999).

\bibitem{annrevnucsci19} J. A. Nolen, and J. P. Schiffer, Annu. Rev. Nucl. Sci. {\bf19}, 471 (1969).
\bibitem{phylett11} K. Okamoto, Phys. Lett. {\bf11}, 150 (1964).
\bibitem{plb277} L. N. Epele, H. Fanchlottl, C. A. Garcia Canal, and G. A. Gonzfilez Sprlnberg,   Phys. Lett {\bf B277}, 33 (1992).
\bibitem{prc63} R. Machleidt, and H. Muther,  Phys. Rev. C {\bf63}, 034005 (2001).



\bibitem{annrevnucsci56} G. A. Miller, A. K. Opper, and E. J. Stephenson, Ann. Rev. Nucl. Part. Sci. {\bf56}, 253 (2006).
\bibitem{prd83} C. E. Wolfe and K. Maltman,   Phys. Rev. D {\bf83}, 077301 (2011).
\bibitem{ppnp39} H. B. O'Connell, B. C. Pearce, A. W. Thomas, and A. G. Williams, Prog. Part. Nucl. Phys. {\bf 39}, 201 (1997). 
\bibitem{plb339} A. K. Dutt-Mazumder, B. Dutt-Roy, and A. Kundu, Phys. Lett. {\bf B399}, 196 (1997).
\bibitem{prc63_015204}  A. K. Dutt-Mazumder, R. Hofmann, and M. Pospelov, Phys. Rev. C {\bf63}, 015204 (2000). 
\bibitem{jpg35} P. Roy, A. K. Dutt-Mazumder, S. Sarkar, and Jan-e Alam,  J. Phys. {\bf G35}, 065106, (2008).

\bibitem{npa249} P. C. McNamee, M. D. Scadron, and S. A. Coon, Nucl. Phys. {\bf A249}, 483 (1975).
\bibitem{npa287} S. A. Coon, and M. D. Scadron, Nucl. Phys. {\bf A287}, 381 (1977). 

\bibitem{prc26} S. A. Coon and M. D. Scadron, Phys. Rev. C {\bf 26}, 562 (19982).
\bibitem{prc48} J. Piekarewicz, Phys. Rev. C {\bf 48},  1555 (1993).


\bibitem{PhyRevC_36} S. A. Coon and R. C. Barrett,  Phys. Rev. C {\bf36}, 2189 (1987).






\bibitem{PhyRevD_62} T. K. Chyi, C. W. Hwang, W.F. Kao, G.L.Lin, K. W. Ng, and J. J. Tseng, Phys. Rev. D {\bf 62}, 105014 (2000).






\bibitem{FBS_12} T. Goldman, J. A. Henderson, and A. W. Thomas, Few-Body Syst. {\bf 12}, 123 (1992).

\bibitem{PhyLettB_317} G. Krein, A. W. Thomas, and A.G. Williams Phys. Lett. {\bf B317} 293 (1993).

\bibitem{PhyLettB_336} H. B. O'Connell, B.C. Pearce, A.W. Thomas, and A.G. Williams, Phys. Lett. {\bf B336} 1 (1994).

\bibitem{PhyRevC_47} J. Piekarewicz and A. G. Williams, Phys. Rev. C {\bf47}, R2462 (1993). 
 
\bibitem{PhyRevC_49} T. Hatsuda, E. M. Henley, Th. Meissner and G. Krein,  Phys. Rev. C {\bf49}, 452 (1994).

  

\bibitem{PhyRevC_52} T. D. Cohen and G. A. Miller, Phys. Rev. C {\bf52}, 3428 (1995).

\bibitem{Aip_412} S. A. Coon, B. H. J. McKellar, and A. A. Rawlinson, AIP Conf. Proc. 412, 368 (1997).
\bibitem{Ppnp_39} H.B. O’Connell, B.C. Pearce, A.W. Thomas, and A.G. Williams, Prog. Part. Nucl. Phys. {\bf 39}, 201 (1997). 
\bibitem{WS_127} G. A. Miller and W. H. T. van Oers, in Symmetries and Fundamental Interactions in Nuclei, edited by W. C. Haxton and 
E. M. Henley (World Scientific, Singapore, 1995), p. 127. 

\bibitem{PhyRevC_80} Wei-Zhou Jiang and Bao-An Li, Phys. Rev. C {\bf80}, 044322 (2009).

\bibitem{PhyRep82}J.~Schwinger,~Phys.~Rev. \textbf{82},664 (1951).

\bibitem{PhyRevC_78} S. Biswas, P. Roy, and A. K. Dutt-Mazumder, Phys. Rev. C {\bf78}, 045207 (2008).

 \end{thebibliography}
\end{document}